# A new diffuse reflector filament for additive manufacturing of 3D printing finely-segmented plastic scintillator

S. Berns,[1, 2, 3] E. Boillat,[1, 2, 3] S. Hugon,[1, 2, 3] A. Boyarintsev,[4] B. Grynyov,[4] N. Karavaeva,[4]
A. Krech,[4] S. Minenko,[4] T. Sibilieva,[4] M. Sibilyev,[4] A. De Roeck,[5] T. Dieminger,[6] U.
Kose,[6] B. Li,[6] A. Rubbia,[6] D. Sgalaberna,[6] T. Weber,[6] J. Wüthrich,[6, *] and X. Zhao[6]

[1]*Haute Ecole Spécialisée de Suisse Occidentale (HES-SO),*
*Route de Moutier 14, CH-2800 Delémont, Switzerland*
[2]*Haute Ecole d'Ingénierie du canton de Vaud (HEIG-VD),*
*Route de Cheseaux 1, CH-1401 Yverdon-les-Bains, Switzerland*
[3]*COMATEC-AddiPole, Technopole de Sainte-Croix,*
*Rue du Progrès 31, CH-1450 Sainte-Croix, Switzerland*
[4]*Institute for Scintillation Materials NAS of Ukraine (ISMA), Nauky ave. 60, 61072 Kharkiv, Ukraine*
[5]*Experimental Physics Department, CERN, Esplanade des Particules 1, 1211 Geneva 23, Switzerland*
[6]*ETH Zurich, Institute for Particle Physics and Astrophysics, CH-8093 Zurich, Switzerland*

This study presents the development and characterization of a novel white reflective filament suitable for additive manufacturing of finely segmented plastic scintillators using 3D printing. The filament is based on polycarbonate (PC) and polymethyl methacrylate (PMMA) polymers loaded with titanium dioxide ($TiO_2$) and polytetrafluoroethylene (PTFE) to enhance reflectivity. A range of filament compositions and thicknesses was evaluated through optical reflection and transmittance measurements. Reflective layers were made by using the Fused Deposition Modeling (FDM) technique. A 3D-segmented plastic scintillator prototype was made with fused injection modeling (FIM) and tested with cosmic rays to assess the light yield and the optical crosstalk. The results demonstrate the feasibility of producing compact and modular 3D-printed scintillator detectors with a performance analogous to standard plastic scintillator detectors, with lower light crosstalk, thus higher light yield, compared to past works, owing to the improved optical properties of the reflector material.

## I. INTRODUCTION

Plastic scintillators (PS) are widely employed in particle detectors due to their fast response and ease of fabrication. They are often deployed in time of flight detectors [1, 2], tonne-to-kilotonne neutrino active detectors [3–5], sampling calorimeters [6], or scintillating optical fibers [7]. Recently, major advancements in the development of novel three-dimensional (3D) granular scintillating detectors for imaging electromagnetic and hadronic showers [8], as well as neutrino interactions [9, 10] have been achieved. Owing to sub-ns response, these detectors are also used for efficient neutron detection with kinetic energy reconstruction by time of flight [11]. Conventionally, they are manufactured using cast polymerization [12], injection molding [13], or extrusion [14, 15] techniques, followed by mechanical post-processing to achieve the desired geometrical precision. However, these methods are labor-intensive and limit the complexity of achievable detector geometries. An attempt to simplify detector fabrication included the development of a prototype of PS cubes glued together, achieving a tolerance of approximately 200 $\mu m$ [16]. However, such a method is not feasible for the production of a single 3D volume of PS cubes, but only of 2D layers. Recent advances in additive manufacturing have introduced new opportunities for producing scintillator elements with high spatial segmentation [17, 18]. In particular, Fused Deposition Modeling (FDM) allows simultaneous printing of scintillation and reflective materials using two extruders, thereby automating the fabrication of finely segmented detectors [19].

Our previous work demonstrated that filaments composed of polystyrene (PST) and polymethyl methacrylate (PMMA) blended with $TiO_2$ could serve as effective reflective components for such applications, [19]. While PST-based reflectors exhibited high reflectivity, they were incompatible with polystyrene scintillators due to material mixing during simultaneous printing [20]. PMMA-based filaments, in contrast, provided a viable solution for manufacturing optically isolated scintillating structures. The light yield was found to be quite uniform among the different cubes of the matrix, and the optical crosstalk was found to be less than 2% for the 3D-printed matrix layer with 1 mm thick white layer, acceptable for applications that require a combined particle tracking and calorimetry [19]. Beyond our work, several developments on 3D printing of plastic scintillator are ongoing on FDM [21], including neutron-sensitive filaments [22], and resins [23–27] also for neutron identification [28–30].

Nevertheless, challenges remained in achieving the required transparency and dimensional accuracy for large-scale detector applications. To overcome this, we introduced a hybrid 3D printing approach, named Fused In-

---

jection Modeling (FIM), which combines the geometric flexibility of FDM with the material quality of injection molding [31, 32]. In this method, a hollow matrix of reflective material is first printed, and then filled with molten scintillator using a heated injector. This imposes stringent thermal requirements on the reflector filament, which must retain its structural integrity at injection temperatures of about 230°C.

In this study, we present the formulation, extrusion, optical characterization, and detector-level validation of several new white reflective filaments based on polycarbonate (PC) and poly methyl methacrylate (PMMA) matrices loaded with titanium dioxide (TiO$_2$) and polytetrafluoroethylene (PTFE). Our goal is to identify optimal material combinations for high-performance 3D-printed scintillator detectors, to improve both the optical properties and, at the same time, achieve a good tolerance with manufacturing without any subtractive process..

## II. REFLECTIVITY AND TRANSMITTANCE MEASUREMENTS

### A. Experimental details

Reflective filaments were fabricated by thermally extruding polymer-additive mixtures using both laboratory and industrial-scale equipment. As the methodology discussed in [20], a Noztek ProHT desktop extruder [33] was used for initial prototyping. For larger-scale production and consistency checks, an industrial Battenfeld extruder and a Noztek Filament Winder 1.0 [34] were utilized (see Fig. 1).

Polymer granules were first mixed with reflective additive powders, specifically TiO$_2$ and PTFE, in a batch mixer to ensure uniform distribution. Then, the mixture was fed into the extruder, and the resulting filament had a diameter of $1.75 \pm 0.1$ mm, suitable for standard FDM 3D printers.

A Creatbot F430 dual-extruder 3D printer [35] was used to fabricate test samples of reflective material for optical characterization. Square samples ($20 \times 20$ mm) with varying thicknesses of 0.2 mm, 0.4 mm, and 1.0 mm were printed using the extruded filaments. The printing temperature was set to 265° for all samples to ensure consistency across material types and layer heights.

The optical properties of the printed samples were measured using a Shimadzu UV-2450 spectrophotometer equipped with an integrating sphere. Measurements of optical reflection $R$ and light transmittance $T$ were performed over the spectral range of 200–800 nm [20]. The instrument's measurement uncertainty was about 0.5%.

Additional characterization of thermal stability and mechanical integrity under printing conditions was carried out to ensure filament suitability for Fused Injection Modeling (FIM), a novel additive manufacturing method optimised for producing optically-isolated segmented highly-transparent cubes of scintillating materi-als where a white empty reflective mold is 3D printed with FDM and filled with melted transparent plastic scintillator with a dedicated nozzle [31, 32].

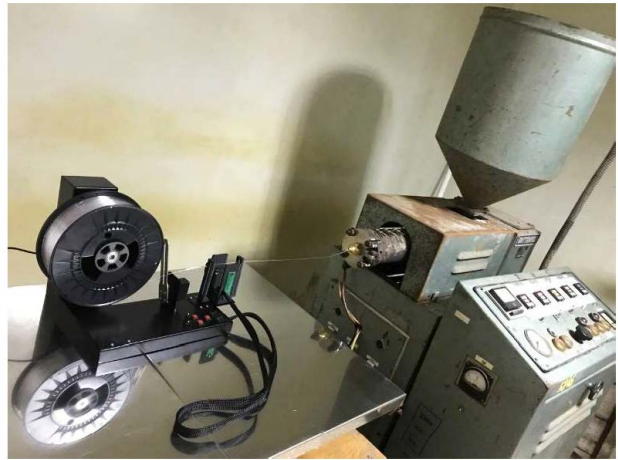

FIG. 1: Photograph of the equipment complex that was used for producing filaments.

## III. RESULTS AND DISCUSSIONS

The selection of an optimal reflective additive is a crucial step in developing a white filament with high optical reflectivity suitable for use in 3D-printed scintillator detectors. The ideal additive must maximize reflectivity at wavelengths in the range between 400 and 500 nm, the characteristic emission peak of polystyrene-based scintillators, while maintaining compatibility with polymer extrusion and printing processes. We initially compared the reflectivity of various powdered reflector materials before incorporating them into polymer matrices.

### 1. Reflectivity of Additive Powders

Pressed powder samples of candidate reflective additives were prepared, and then the optical reflection measurements were performed to evaluate their performance independently of polymer matrices. The results are summarized in Fig. 2.

As shown, in the region of scintillation polystyrene luminescence, at 429 nm, pure polytetrafluoroethylene (PTFE) showed the highest reflectivity, with measured values exceeding 100%, defined from the reference measurement. The optical reflection $R$ was measured on the spectrophotometer compared to the BaSO$_4$ in the integrating sphere. The reflection value of the reference sample BaSO$_4$ (in the integrating sphere) was taken as 100%. Titanium dioxide (TiO$_2$) also demonstrated good performance, with a reflectance of 94.64%. However, combining TiO$_2$ and PTFE in a 2:1 weight ratio did not yield a significant enhancement over the individual components, achieving a reflectance of 94.3%. Based on these results, we selected specific concentrations of $TiO_2$ and PTFE for filament production.

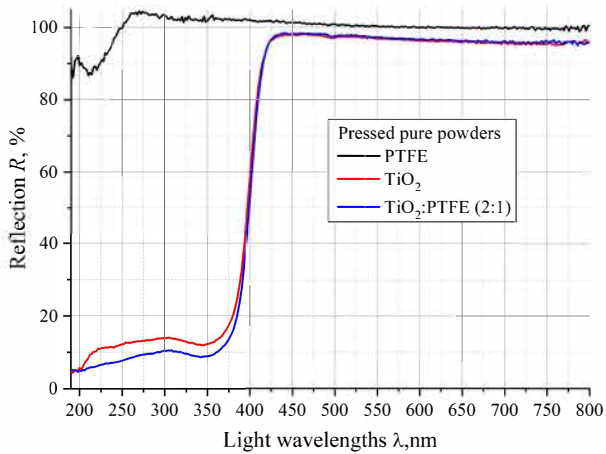

FIG. 2: Results of optical reflection measurement of powders.

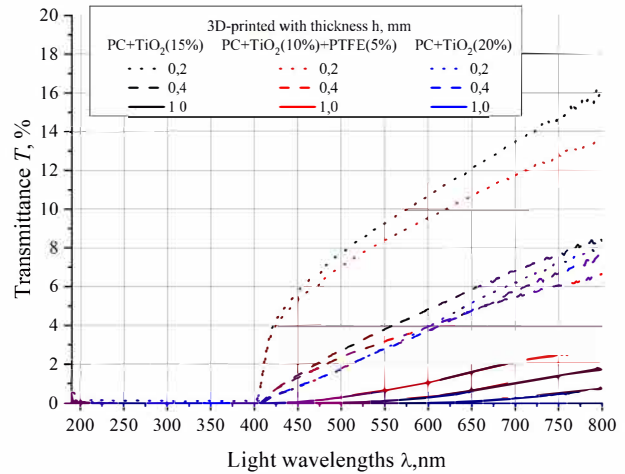

FIG. 4: Comparison of light transmittance in 3D-printed samples PC with varying concentrations of reflective additives depending on sample thickness.

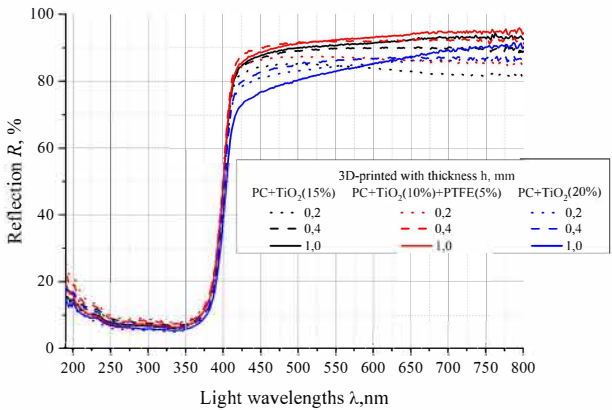

FIG. 3: Comparison of light reflection in 3D-printed samples of polycarbonate (PC) with varying concentrations of reflective additives depending on sample thickness.

#### 2. 3D-Printed PC-based Filament Samples

Following the powder reflectivity tests, effective filaments were fabricated using polycarbonate (PC) as the base polymer, combined with various concentrations of $TiO_2$ and PTFE as reflective additives. These filaments were used to 3D-print samples of different thicknesses (0.2 mm, 0.4 mm, and 1.0 mm), which were then optically characterized. As shown in Fig. 3, the formulation containing 10% $TiO_2$ and 5% PTFE demonstrated the highest reflectivity among all PC-based samples, especially in the 410–450 nm wavelength range. Notably, even the 0.2 mm thick sample achieved reflectivity levels comparable to those of thicker (0.4 mm and 1.0 mm) samples, with the same or higher additive content. However, increasing $TiO_2$ concentration to 20% led to decreased performance. It was attributed to the increased brittleness and porosity of the filaments, which negatively impacted the surface smoothness and the internal scattering.

#### 3. Light Transmittance of PC-based Samples

The light transmittance measurement of different samples with various thicknesses is presented in Fig. 4. The reflectivity reaches the few percent level already with a thickness of 0.4 mm. Although 0.2 mm-thick samples also provide satisfactory performance, their reflectance is found to be 2–3% worse than that of the 0.4 mm samples. A summary of results is presented in Table I. Samples composed of PC with $TiO_2$ (10%) and PTFE (5%) exhibited lower light transmittance compared to PC + $TiO_2$ (15%) samples. The values of light transmittance of the samples with thicknesses 0.2, 0.4, and 1.0 mm at the wavelength of 420 nm for comparison are shown in Table I. As expected, the lowest light transmittance is observed in 1.0 mm-thick samples due to their greater optical path length. However, 0.4 mm-thick samples already offer excellent opacity, with transmittance in the range of 0.3–0.6%, making them highly suitable for use as reflective material in fine-granularity optically segmented detectors. These results establish the PC + 10% $TiO_2$ + 5% PTFE formulation, printed at a 0.4 mm thickness, as the optimal configuration for reflective 3D-printed structures. At this thickness, and at 420 nm wavelength the highest reflectivity of (86.22%) and a light transmittance of $T = 0.55\%$. The minimum effective thickness for functional reflectivity is 0.2 mm, yielding a reflectivity of $R = 83.35\%$ and a light transmittance of $T = 3.6\%$. These results confirm that 0.4 mm offers the best balance between high reflectivity and low light transmittance, while maintaining mechanical stability and printability.

| Sample Composition | Reflectivity R (%) | | | Transmittance T (%) | | |
|---|---|---|---|---|---|---|
| | 0.2 mm | 0.4 mm | 1.0 mm | 0.2 mm | 0.4 mm | 1.0 mm |
| PC + TiO$_2$ (15%) | 81.38 | 82.96 | 83.36 | 3.81 | 0.55 | 0 |
| PC + TiO$_2$ (10%) + PTFE (5%) | 83.35 | 86.22 | 84.47 | 3.60 | 0.55 | 0.01 |
| PC + TiO$_2$ (20%) | 77.23 | 78.55 | 71.70 | 0.35 | 0.32 | 0 |
| PC + PTFE Nanovia [20] | – | – | 68.69 | – | – | 18.67 |
| PMMA + TiO$_2$ (10%) + PTFE (5%) | 84.73 | 90.53 | 92.07 | 10.36 | 3.69 | 0.1 |
| PMMA + TiO$_2$ (15%) | 91.95 | 92.53 | 91.1 | 1.57 | 1.69 | 0.01 |

TABLE I: Comparison of reflectivity $R$ and transmittance $T$ for 3D-printed samples. $\lambda = 420$ nm (%). Each sample was printed at the indicated thickness and characterized optically.

### 4. Comparison with PMMA-based Filaments

To assess the potential of PMMA as an alternative base polymer, the same additive mixture (10% TiO$_2$ and 5% PTFE) was introduced into PMMA filaments. Samples were printed in various thicknesses and characterized under identical conditions. As shown in Fig. 5, PMMA-based samples exhibited both higher reflectivity and higher transmittance compared to PC-based samples. Specifically, at 420 nm, the reflectivity of 0.4 mm PMMA samples reached 90.53%, exceeding that of PC samples. However, this came at the cost of increased light transmittance (3.69% for PMMA vs 0.55% for PC at 0.4 mm thickness), which may result in reduced optical isolation between adjacent scintillator voxels in a detector. A detailed comparison of reflectivity and transmittance values for both materials at 420 nm is provided in Table I.

Given that PMMA-based samples showed higher reflectivity than PC-based ones, with $R = 90.53\%$ at $h = 0.4$ mm compared to $R = 86.22\%$ for PC at the same thickness, PMMA can be considered as a promising base material for reflective filaments. However, this advantage in reflectivity is accompanied by significantly higher light transmittance, which may compromise optical isolation in segmented detector applications. For example, at a thickness of 0.2 mm, PMMA exhibits a transmittance of $T = 10.36\%$, compared to only 3.60% for PC; similarly, at 0.4 mm, the transmittance is 3.69% for PMMA, whereas PC maintains a much lower value of 0.55%.

To investigate whether PTFE is necessary in the PMMA formulation, we produced a new filament containing PMMA with 15% TiO$_2$ only, excluding PTFE. It should be noted that in PMMA we have more light absorption. Samples were 3D-printed and characterized under the same conditions. The measured reflection and transmittance values are presented in Fig. 6 and given in Table I. PC-based samples exhibit higher light transmittance than PMMA-based samples, but a worse light reflection $R$=90.53% at $h$=0.4 mm for PMMA-based samples and $R$=86.22% at the same thickness for PC-based samples.

These results indicate that PMMA loaded with 15% TiO$_2$ is a highly promising candidate for reflective applications, delivering exceptional optical performance even at minimal thickness. However, it should be noted that the softening temperature of PMMA is lower than that of PC. The typical glass temperatures for PMMA range between 85 °C and 105 °C, while for PC it is approximately ~140 °C.

## IV. SUPERLAYER MEASUREMENT

To characterize the performance of the developed reflector filament on real detector prototypes, multiple segmented plastic scintillator particle detector prototypes, called the "SuperLayers", were manufactured with the FIM technique.

Each SuperLayer consists of $4 \times 4 \times 1$ plastic scintillator cubes, each with a dimension of $1 \times 1 \times 1$ cm$^2$. All the six faces of each cube were surrounded by 1 mm thick reflective walls fabricated using the newly developed 3D-printed while filaments, aiming to ensure optical isolation and maximize light collection. Two SuperLayers were manufactured, each utilizing different reflector filament recipes: PMMA + 10% TiO$_2$ + 5% PTFE, and PC + 10% TiO$_2$ + 5% PTFE. A fully 3D-printed particle detector consisting of $5 \times 5 \times 5$ scintillator cubes (the "SuperCube", whose production and performance have been reported in [31, 32]) was adopted as the reference and measured together with the SuperLayers. The scintillation light produced by charged particles traversing the cubes was collected via two wavelength-shifting (WLS) fibers inserted along orthogonal axes of each cube. Each SuperLayer was read out by eight WLS fiber channels, coupled to Hamamatsu S13360-1325 Multi-Pixel Photon Counters (MPPCs). As front-end board (FEB), the CAEN

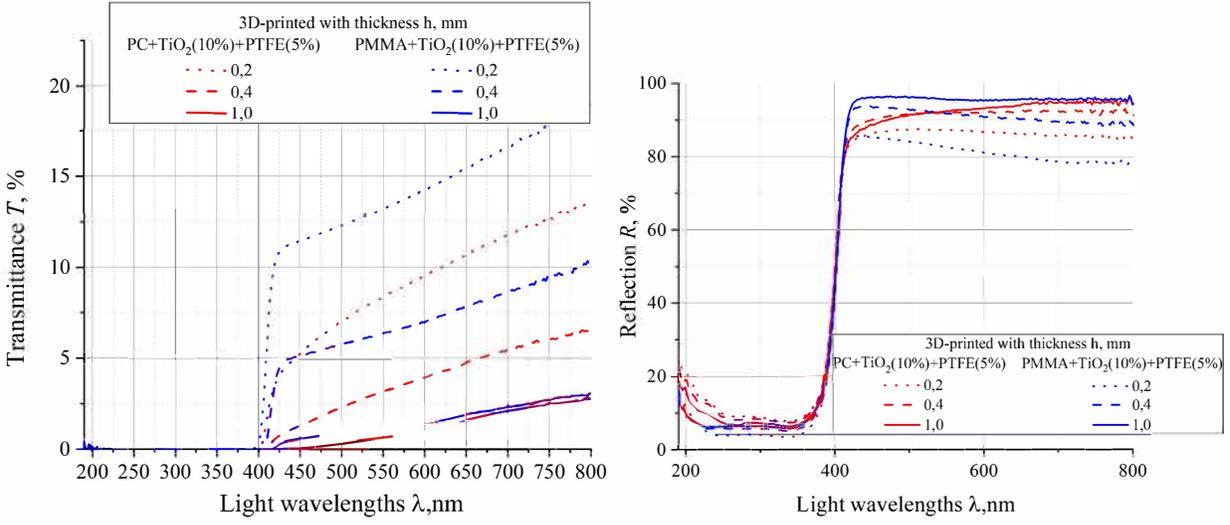

FIG. 5: Comparison of light transmittance $T$ (left) and light reflection $R$ (right) in 3D-printed PC and PMMA samples with identical concentrations of reflective additives depending on sample thickness.

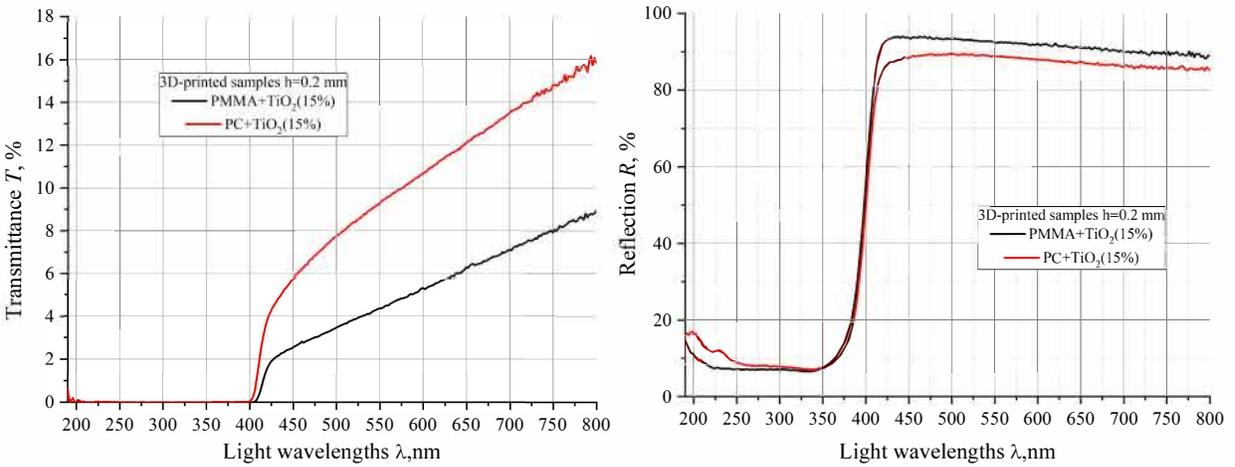

FIG. 6: Comparison of light transmittance $T$ (left) and light reflection $R$ (right) in 3D-printed PC and PMMA samples with identical concentrations of reflective additives (15% $TiO_2$).

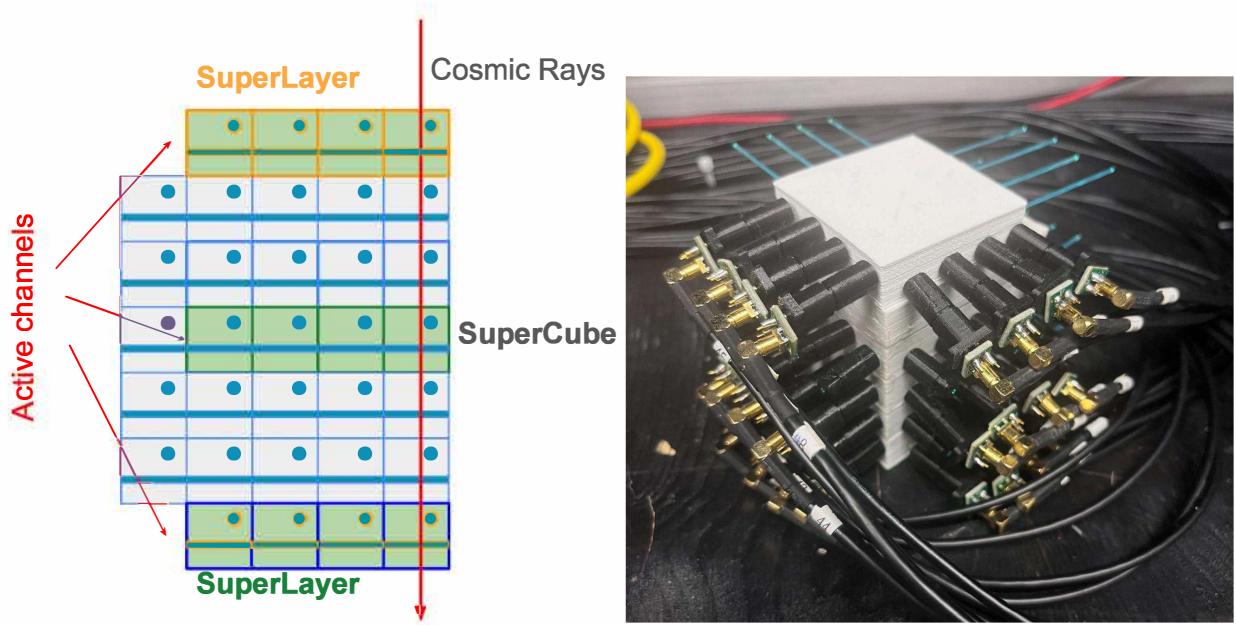

FIG. 7: Left: A schematic showing the measurement setup. Right: One SuperLayer stacked on top of the SuperCube with all fiber inserted and MPPC coupled, ready for measurement.

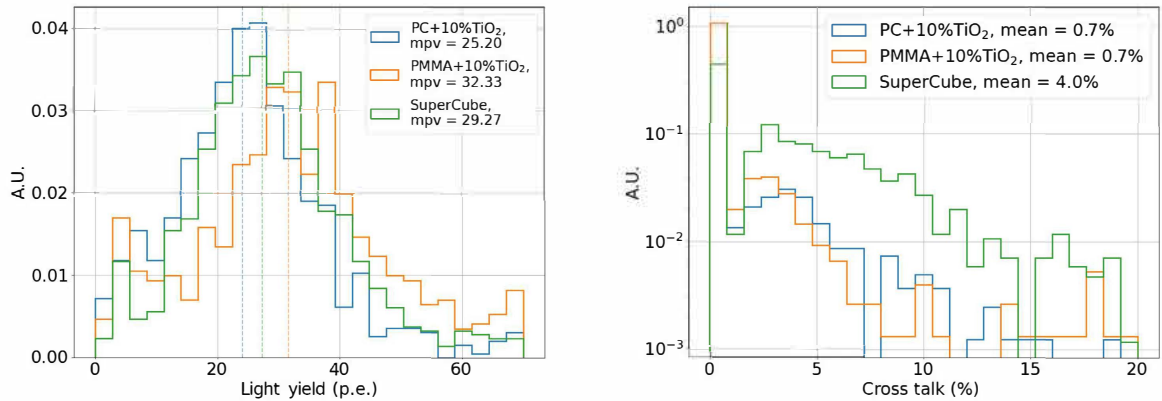

FIG. 8: Left: The single channel light yield distributions measured with three different cube layers. Right: The cube-to-cube light crosstalk distributions measured with three different cube layers.

DT5202 [36], was used to digitize the analog signals from the MPPCs. The concept of the measurement is illustrated in the left panel of Fig. 7, while an example of the fully instrumented setup can be seen in the right panel of Fig. 7. Two SuperLayers were stacked on top and bottom of the SuperCube to ensure geometrical symmetry. The 4 × 4 cubes in the center layer of the SuperCube were measured as a baseline reference. Vertical going cosmic rays were selected by requiring the track passing through three cube layers at the same position. The experimental setup allows to compare our old 3D printed prototype with the new one with improved optical segmentation.

The channel light yield and cube-to-cube light crosstalk were measured to characterize the perfor-mance of each prototype. The results are summarized in Fig. 8. The left panel shows the channel light yield measured with different layers. The peak of each distribution corresponds to the channel-wise detector response to a characterizing energy deposition of $\sim 1.8$ MeV/cm by cosmic rays, typically minimum-ionizing particles (MIPs). As expected, the Super-Layer with the PMMA reflector shows the highest light yield of about 32 p.e./MIP/channel, while the prototype with the PC reflector has a lower light yield of about 25 p.e./MIP/channel compared to the light yield measured with the SuperCube (about 29 p.e./MIP/channel).

The crosstalk was measured as the ratio of the light yield measured by two parallel channels, reading the main cube traversed by the particle track and an adjacent cube, respectively. A threshold of 0.7 p.e. was applied to the crosstalk channels. For all the read out signals below 0.7 p.e.(mostly due to the fluctuation of the electronic pedestal), zero crosstalk was assumed, while all the signals above the threshold were used to calculate the crosstalk ratio. The final distributions are shown in the right panel of Fig. 8. The peaks at zero correspond to the events where the crosstalk is below the threshold. Both the prototypes using PMMA or PC based reflectors show low crosstalk with an average of 0.7%. As a comparison, the SuperCube, using the commercial white reflector filament shows a mean crosstalk of 4% per face.

## V. CONCLUSIONS

In this work, we developed and characterized new white reflective filaments for use in 3D printing finely segmented plastic scintillator detectors. Two base polymers, polycarbonate (PC) and polymethyl methacrylate (PMMA), were evaluated in combination with reflective additives such as titanium dioxide ($TiO_2$) and polytetrafluoroethylene (PTFE).

Optical measurements demonstrated that for PC-based filaments, the optimal formulation consisted of 10% $TiO_2$ + 5% PTFE, yielding a reflectivity of 86.2% at 420 nm with low transmittance ( 0.55%) at 0.4 mm thickness. For PMMA-based filaments, $TiO_2$ alone (at 15%) was sufficient to achieve even higher reflectivity, reaching

91.95% at 0.2 mm thickness. The addition of PTFE to PMMA did not improve the performance, likely due to the inherently lower light absorption and higher transparency of the PMMA matrix. In such optically transparent environments, multi-component formulations may not yield additional benefit, although this outcome may also depend on specific 3D printing conditions and processing parameters.

To evaluate the performance of these materials in realistic detector conditions, segmented scintillator prototypes called SuperLayers were constructed using the developed filaments and tested with vertical cosmic-ray muons. The SuperLayer built with the PMMA-based reflector achieved the highest light yield, measuring approximately 32 photoelectrons per MIP per channel, outperforming both the PC-based reflector prototype (25 p.e./MIP/channel) and the SuperCube reference detector (29 p.e./MIP/channel), manufactured with the most reflective commercial filament that could be found..

Importantly, both PMMA- and PC-based SuperLayers exhibited excellent optical isolation, with average cube-to-cube light crosstalk as low as 0.7%, significantly lower than the 4% observed in detectors using commercial white reflector filaments. These results confirm that the newly developed 3D-printed reflective filaments are not only compatible with fine voxelized detector geometries, but also provide enhanced light collection and superior suppression of optical crosstalk—key requirements for high-resolution calorimetry and particle tracking applications.

## VI. ACKNOWLEDGEMENTS

This work was supported by the joint grant IZURZ2_224819 of the Swiss National Science Foundation (SNSF) and the National Research Foundation of Ukraine (NRFU). This work was also supported by the SNSF grant PCEFP2_203261.

## VII. AUTHOR INFORMATION

### A. Authors and Affiliations

**A. Krech, T. Sibilieva, A. Boyarintsev, B. Grynyov, N. Karavaeva, S. Minenko, M. Sibilyev** - Institute for Scintillation Materials NAS of Ukraine;

**T. Dieminger, U. Kose, B. Li, A. Rubbia, D. Sgalaberna, T. Weber, X. Zhao** - ETH Zurich, Institute for Particle Physics and Astrophysics;

**S. Berns, E. Boillat, S. Hugon** - Haute Ecole Spécialisée de Suisse Occidentale (HES-SO), Haute Ecole d'Ingénierie du canton de Vaud (HEIG-VD), COMATEC-AddiPole, Technopole de Sainte-Croix;

**A. De Roeck** - Experimental Physics Department, CERN.

### B. Corresponding author

Anton Krech (antonkrech@gmail.com)

## VIII. ETHICS DECLARATIONS

### A. Competing interests

The authors declare no competing interests.